\documentclass[amsmath]{llncs}
\usepackage{graphicx}	
\usepackage{amsmath}
\usepackage{bm}

\usepackage[dvipsnames]{xcolor}
\definecolor{darkgreen}{hsb}{.333,1,.5}
\definecolor{darkblue}{hsb}{.667,1,.75}
\definecolor{darkbrown}{hsb}{0,.5,.5}
\usepackage{hyperref}	
\hypersetup{
    colorlinks=true,
    linkcolor=darkbrown,	
    urlcolor=darkblue,		
    citecolor=darkgreen,	
}
\usepackage{caption}	
\newcommand{\ee}{e}		

\newcommand{\qast}{\star}
\newcommand{\qi}{i}
\newcommand{\qj}{j}
\newcommand{\qk}{k}
\newcommand{\reverse}[1]{\overline{#1}}	

\newcommand{\bgamma}{\bm{\gamma}}

\newcommand{\bn}{\bm{n}}
\newcommand{\bp}{\bm{p}}

\newcommand{\brr}{\bm{r}}
\newcommand{\bv}{\bm{v}}
\newcommand{\btheta}{\bm{\theta}}

\begin{document}

\title{The rules of 4-dimensional perspective: How to implement Lorentz transformations in relativistic visualization}
\titlerunning{The rules of 4-dimensional perspective}

\author{Andrew J. S. Hamilton \orcidID{0000-0002-3816-5973}}
\institute{%
JILA and Dept.\ Astrophysical \& Planetary Sciences,
Box 440, U. Colorado Boulder, CO 80309, USA
\email{Andrew.Hamilton@colorado.edu}
\url{http://jila.colorado.edu/~ajsh/}
}


\newcommand{\beamfig}{
    \begin{figure}[tb!]
    \begin{center}
    \leavevmode
    \includegraphics[scale=.9]{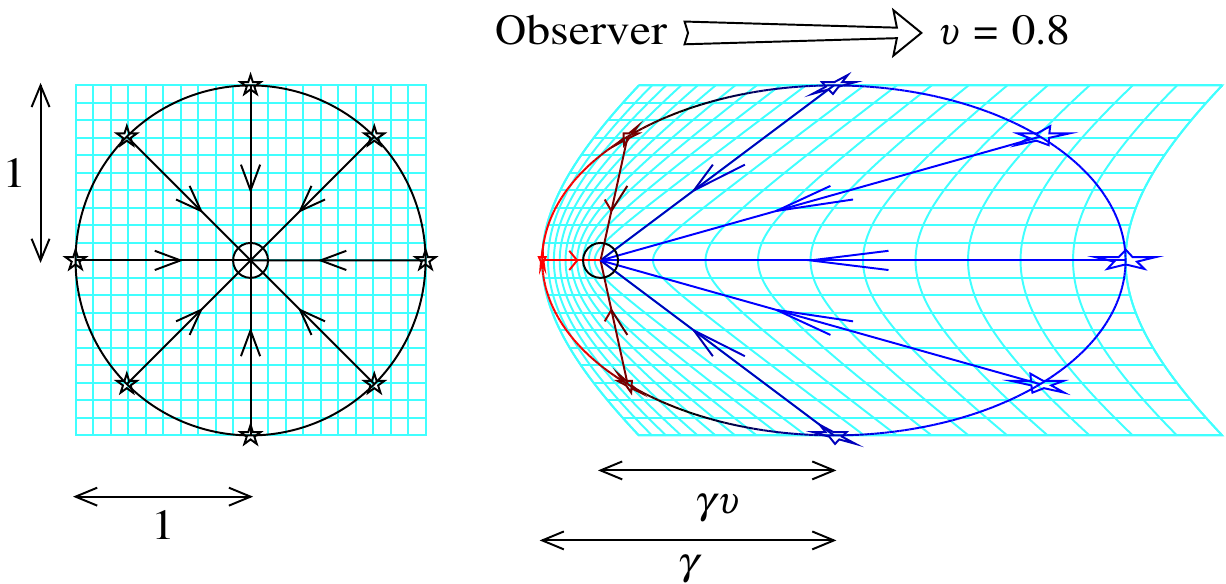}
    \caption[The rules of 4-dimensional perspective]{
    \label{beam}
The rules of 4-dimensional perspective.
In special relativity,
the scene seen by an observer moving through the scene (right)
is relativistically beamed
compared to the scene seen by an observer at rest relative to the scene (left).
On the left,
the observer
at the center of the circle is at rest
relative to the surrounding scene.
On the right,
the observer
is moving to the right through
the same scene at $v = 0.8$ times the speed of light.
The scene is distorted into a celestial ellipsoid with the observer
displaced to its focus.
The arrowed lines represent energy-momenta of photons.
The length of an arrowed line
is proportional to the perceived energy of the photon.
The scene ahead of the moving observer appears concentrated, blueshifted,
and farther away,
while the scene behind appears expanded, redshifted, and closer.
    }
    \end{center}
    \end{figure}
}

\newcommand{\spherefig}{
    \begin{figure}[t!]
    \begin{center}
    \leavevmode
    \includegraphics[scale=.1]{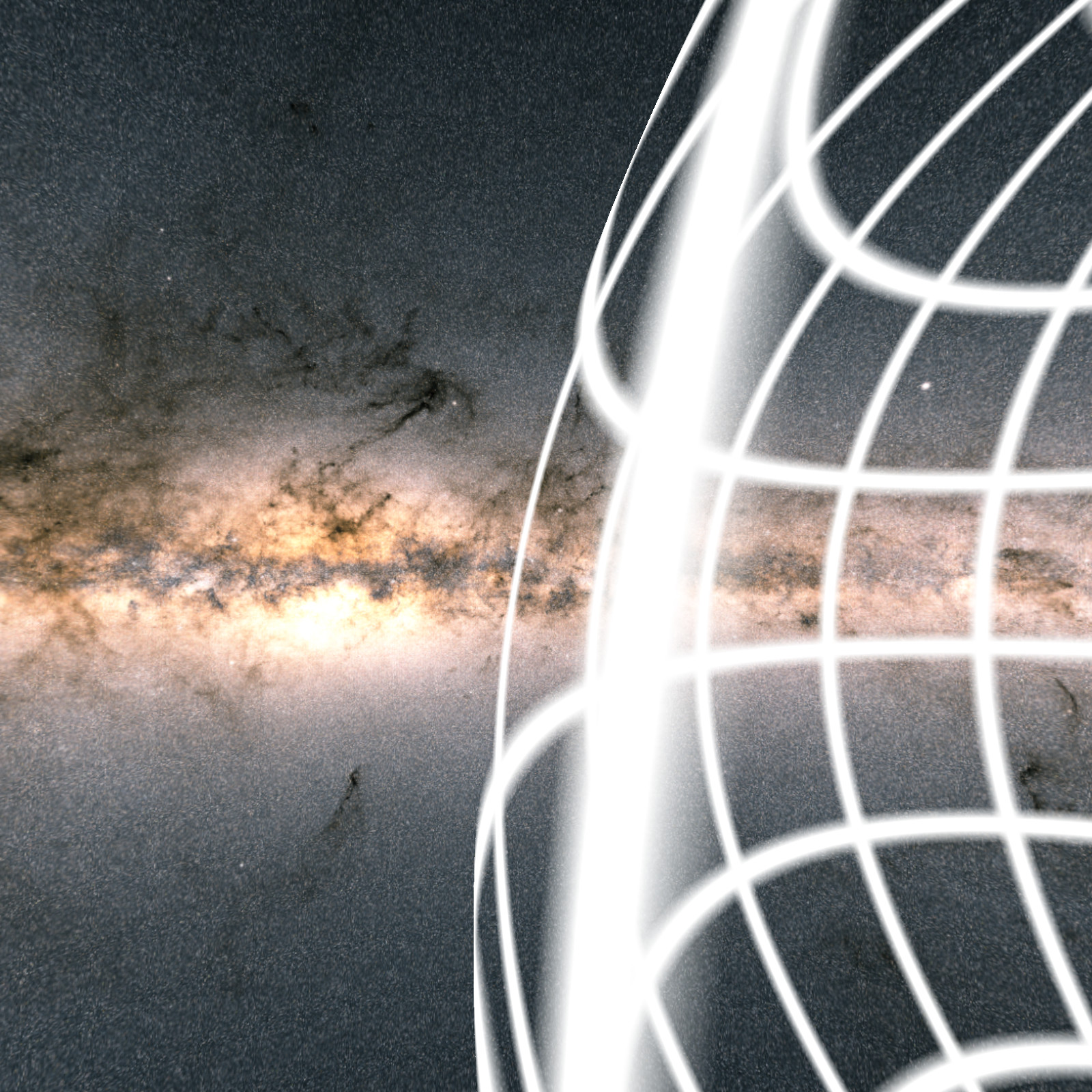}
    \hspace{.5em}
    \includegraphics[scale=.1]{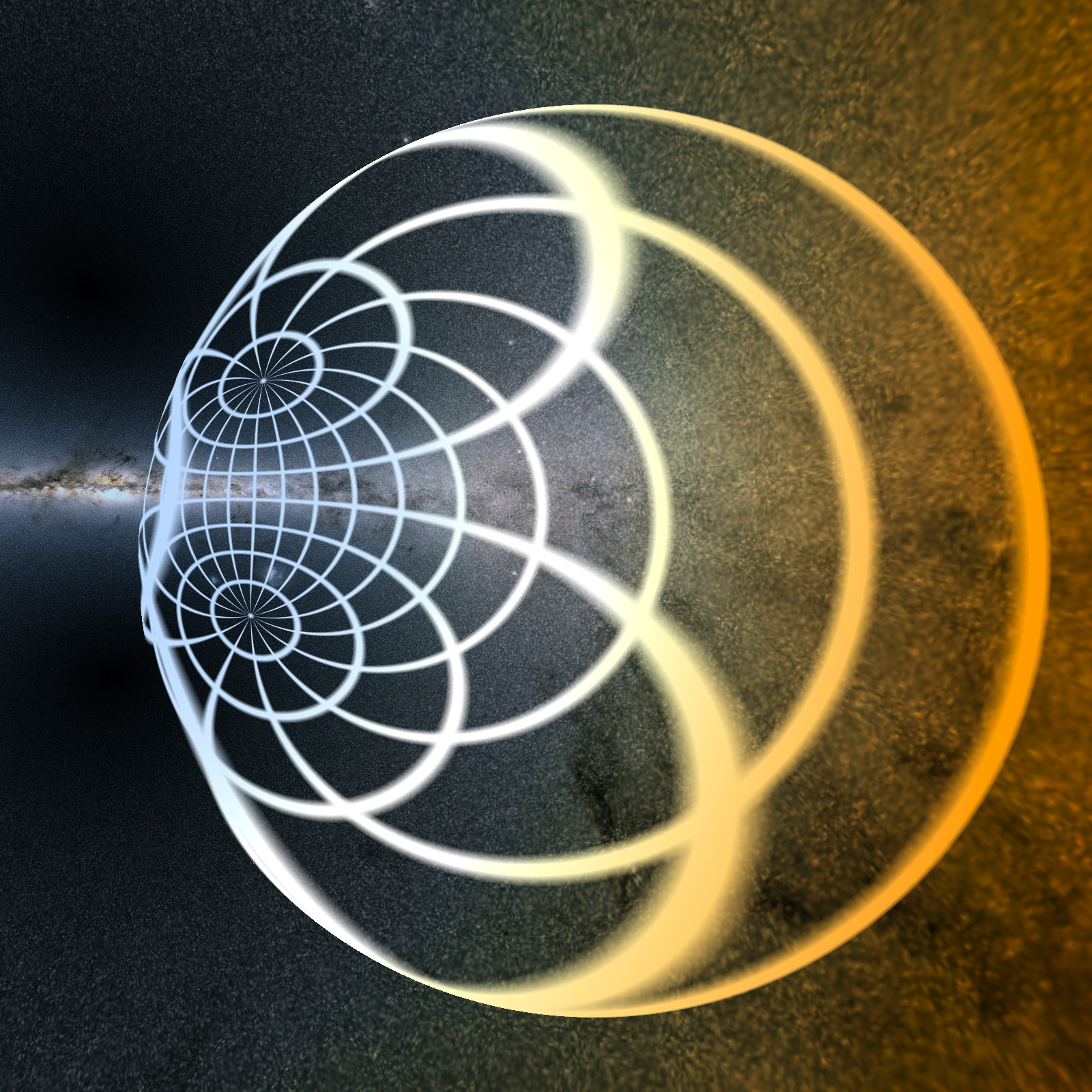}
    \caption[A sphere seen at relativistic speed]{
    \label{sphere}
Passing by a sphere at $0.97$ of the speed of light
($\gamma v = 4$),
in the general direction of the center of our Galaxy, the Milky Way
(left) undistorted, (right) relativistically beamed.
The sphere is painted with lines of latitude and longitude.
The background is an image of the Milky Way from Gaia Data Release~3
\cite{GaiaDR3}.
In the undistorted (left) view,
most of the sphere is behind the observer;
relativistic aberration (right) brings the sphere behind into view.
The sphere and background
are colored with appropriately blue- and red-shifted blackbody colors;
the unredshifted color temperature is $5{,}780 \, \textrm{K}$,
the color of the Sun.
The field of view is $105^\circ$ across the diagonal.
The visualization was made using the
Black Hole Flight Simulator software
\cite{BHFS}.
    }
    \end{center}
    \end{figure}
}

\maketitle


\begin{abstract}
This paper presents a pedagogical introduction to the issue
of how to implement Lorentz transformations in relativistic visualization.
The most efficient approach is to use
the even geometric algebra in 3+1 spacetime dimensions,
or equivalently complex quaternions,
which are fast, compact, and robust,
and straightforward to compose, interpolate, and spline.
The approach has been incorporated into the Black Hole Flight Simulator,
an interactive general relativistic ray-tracing program
developed by the author.
\end{abstract}

\section{Introduction}

Einstein's theory of Special Relativity revolutionized space and time,
uniting them from the separate 3-dimensional and 1-dimensional entities
familiar to everyday experience,
into an inextricably entwined 4-dimensional spacetime.
The consequences of special relativity are not evident to everyday experience
because we move through our surroundings
at only a tiny fraction of the speed of light.
But if you moved around at near the speed of light,
then you would very much notice
\cite{Kraus:2007,Kraus:2021,Sherin:2016}.

The positions of objects in 3 dimensions are described by
3-dimensional vectors, or 3-vectors for short.
Of primary importance in 3d visualization software is
the ability to rotate 3-vectors rapidly and robustly.
A rotation in 3 dimensions decomposes into rotations about
each of 3 directions.
It is well known that the best way to program rotations
is using quaternions,
invented by William Rowan Hamilton in 1843
\cite{ODonnell:1983}.
The algebra of quaternions is precisely that of the even Clifford algebra,
or geometric algebra, in 3 spatial dimensions.

In relativistic visualization,
the perceived position of an object is described by a 4-dimensional vector,
a 4-vector.
The analog of rotations of a 4-vector are Lorentz transformations
\cite{Lorentz:1904}.
To the usual 3 spatial rotations of space,
Lorentz transformations adjoin 3 rotations between space and time,
also known as boosts,
a total of 6 spacetime rotations.
The 3 boosts represent a change of the observer's velocity
along each of the 3 spatial directions.
To someone moving through a scene at much less than the speed of light,
these boosts have a negligible effect on the appearance of a scene.
But if you move at near the speed of light,
then the boosts are very evident.

The purpose of this paper is to present a pedagogical introduction
to implementing Lorentz transformations in computer visualizations.
The paper starts, \S\ref{4drules-sec},
by presenting a conceptual picture of how a scene appears distorted
to an observer who moves through it at relativistic speed
\cite{4dperspective}.
Then \S\ref{implementlorentz-sec}
describes how to implement Lorentz transformations using complex quaternions,
also called biquaternions by William Rowan Hamilton.
The equivalence of the Lorentz group to complex quaternions
was discovered by Silberstein in 1912 \cite{Silberstein:1912},
and remarked by Dirac \cite{Dirac:1944};
see \cite{Berry:2021} for further references.
The algebra of complex quaternions is precisely that of
the even geometric algebra in 3+1 spacetime dimensions.

In relativity,
time behaves mathematically as if it were an imaginary spatial dimension.
Lorentz transformations rotate spatial and time dimensions among each other.
In relativity it is extremely useful to use ``natural'' units
in which the speed of light is one, $c = 1$,
since then space and time are measured in the same unit.

\section{The rules of 4-dimensional perspective}
\label{4drules-sec}

The conceptual picture presented in this section
is from the author's website \cite{4dperspective},
where animations may be found.

In 3-dimensional perspective, two ideas are fundamental:
\begin{itemize}
\item
A straight line in 3 dimensions remains a straight line in perspective;
\item
Parallel lines meet at a vanishing point.
\end{itemize}
\vspace{2pt}
%

\beamfig

The rules that govern the appearance of a scene
when you move through it at near the speed of light can be called
the rules of 4-dimensional perspective.
These rules can be grasped,
much like the rules of 3-dimensional perspective,
without needing to understand intricate mathematics.
The rules of 4-dimensional perspective
can be summarized as follows:
\begin{itemize}
\item
Paint the scene at rest on the surface of a celestial sphere;
\item
Stretch the celestial sphere by Lorentz factor $\gamma$
along the direction of motion into a celestial ellipsoid,
and displace the observer to a focus of the ellipsoid;
\item
Adjust the brightness, color, and clock speed at any point on the ellipsoid
in proportion to the length of the radius between the point and the observer.
\end{itemize}
\vspace{2pt}

Figure~\ref{beam} illustrates the rules of 4-dimensional perspective.
On the left, you are at rest relative to the scene.
Imagine painting the scene on a celestial sphere around you.
The arrows represent the directions of lightrays (photons)
from the scene on the celestial sphere to you at the center.
Technically, these arrows represent the energy-momentum 4-vectors
of the lightrays that you see, \S\ref{4vec-sec}.

On the right in Figure~\ref{beam},
you are moving to the right through the scene,
at velocity $v = 0.8$ times the speed of light.
The celestial sphere is stretched
along the direction of your motion
by the Lorentz gamma-factor $\gamma = 1/\sqrt{1 - 0.8^2} = 5/3$
into a celestial ellipsoid.
You, the observer, are not at the center of the ellipsoid,
but rather at one of its foci (the left one, if you are moving to the right).
The focus of the celestial ellipsoid, where you the observer are,
is displaced from center by $\gamma v = 4/3$.
The scene appears relativistically aberrated,
which is to say concentrated ahead of you,
and expanded behind you.

The lengths of the arrows in Figure~\ref{beam}
are proportional to the energies, or frequencies, of the photons that you see.
When you are moving through the scene at near light speed,
the arrows ahead of you,
in your direction of motion,
are longer than at rest,
so you see the photons blue-shifted, increased in energy,
increased in frequency.
Conversely, the arrows behind you are shorter than at rest,
so you see the photons red-shifted, decreased in energy, decreased in frequency.
Since photons are good clocks,
the change in photon frequency
also tells you how fast or slow clocks attached to the scene appear to you to run.

\spherefig

The following table summarizes the four effects of relativistic beaming
on the appearance of a scene ahead of you and behind you
as you move through it at near the speed of light:
\begin{center}
  \begin{tabular}{l@{\quad}l@{\quad}l}
  Effect & Ahead & Behind \\
  \hline
  Aberration & Concentrated & Expanded \\
  Color & Blueshifted & Redshifted \\
  Brightness & Brighter & Dimmer \\
  Time & Speeded up & Slowed down \\
  \hline
  \end{tabular}
\end{center}

Figure~\ref{sphere}
shows a visualization of passing by a sphere at near the speed of light,
in this case $v = 0.97$
(in natural units $c = 1$).
The visualization was made using the Black Hole Flight Simulator software
\cite{BHFS}.
The visualization illustrates how relativistic aberration distorts
the appearance of the sphere,
concentrating it and blueshifting it towards the direction the oberver
is moving, and expanding and redshifting it behind.

A feature of relativistic aberration is that circles transform to circles,
and angles are preserved.
These facts were first pointed out by
\cite{Penrose:1959vz} and \cite{Terrell:1959},
prior to which it had been widely thought that
circles would appear Lorentz-contracted and therefore squashed.
Lines of latitude and longitude on the surface of a sphere are circles,
and they intersect at right angles.
Figure~\ref{sphere} illustrates how relativistic aberration
preserves these properties.
The outline of a sphere is itself a circle,
and the outline
of a relativistically aberrated sphere remains circular.

\section{How to implement Lorentz transformations on a computer}
\label{implementlorentz-sec}

The advantages of quaternions
for implementing spatial rotations
are well-known to 3d game programmers.
Compared to standard rotation matrices,
quaternions offer increased speed (fewer multiplications)
and require less storage,
and their algebraic properties
simplify composition, interpolation, and splining.

Complex quaternions retain
similar advantages
for implementing Lorentz transformations.
They are fast, compact,
and straightforward to interpolate or spline.
Moreover,
since complex quaternions contain real quaternions,
Lorentz transformations can be implemented as an extension of spatial rotations
in 3d programs that use quaternions to implement spatial rotations.

At a deeper level,
the elegant properties of quaternions and complex quaternions
can be traced to the fact that they are the even geometric subalgebras
in respectively 3 spatial dimensions and 3+1 spacetime dimensions.

\subsection{Real quaternions}

It is useful to start by reviewing basic definitions and properties
of ordinary (real) quaternions.
Many of these properties carry over to complex quaternions.

A quaternion $q$ is a kind of souped-up complex number,
\begin{equation}
\label{qdef}
  q = w + \qi x + \qj y + \qk z
  \ ,
\end{equation}
where
$w,x,y,z$ are real numbers,
and the three imaginary numbers
$\qi, \qj, \qk$,
are defined to satisfy
\begin{equation}
\label{idef}
  \qi^2 = \qj^2 = \qk^2 =
  - \qi \qj \qk
  = - 1
  \ .
\end{equation}
The convention $\qi \qj \qk = 1$ in the definition~(\ref{idef})
allows $\qi$, $\qj$, $\qk$ to be identified as bivectors~(\ref{basis3e})
of the 3d geometric algebra,
but is opposite to the traditional definition
$ijk = -1$
famously carved by William Rowan Hamilton in the stone of Brougham Bridge
while walking with his wife along the Royal Canal to Dublin on 16 October 1843
\cite{ODonnell:1983}.
A consequence of equations~(\ref{idef})
is that each pair of imaginary numbers anticommutes:
\begin{equation}
\label{ij}
  \qi\qj = -\qj\qi = - \qk
  \ , \quad
  \qj\qk = -\qk\qj = - \qi
  \ , \quad
  \qk\qi = -\qi\qk = - \qj
  \ .
\end{equation}
Quaternions are distributive and associative, but not commutative.
Similarly to complex numbers,
two quaternions are multiplied by multiplying and combining their components.

The algebra of quaternions can be recognized as that of the even
geometric algebra in 3 spatial dimensions.
The 3d geometric algebra is generated by the
$1 + 3 + 3 + 1 = 8 = 2^3$ orthornormal basis multivectors
\begin{equation}
\label{basis3e}
  \begin{array}{c}
    1 \, , \\
    \mbox{1 scalar}
  \end{array}
  \ \ 
  \begin{array}{c}
    \bgamma_x \, , \  \bgamma_y \, , \  \bgamma_z \, , \\
    \mbox{3 vectors}
  \end{array}
  \ \ 
  \begin{array}{c}
    I_3 \bgamma_x \, , \  I_3 \bgamma_y \, , \  I_3 \bgamma_z \, , \\
    \mbox{3 bivectors (pseudovectors)}
  \end{array}
  \ \ 
  \begin{array}{c}
    I_3 \equiv \bgamma_x \bgamma_y \bgamma_z \, . \\
    \mbox{1 pseudoscalar}
  \end{array}
\end{equation}
The quaternionic imaginaries $\qi$, $\qj$, $\qk$
are just the basis bivectors $I_3 \bgamma_a$ of the 3d geometric algebra.

A quaternion $q$ can be stored as a 4-component object
\begin{equation}
\label{quat}
  q
  =
  \bigl\{
  \begin{array}{c@{\ \ }c@{\ \ }c@{\ \ }c}
  w & x & y & z \\
  \end{array}
  \bigr\}
  \ .
\end{equation}
The convention to put the scalar component $w$ in the first position
is standard in mathematics and physics.

A basic operation on quaternions is quaternionic conjugation,
also called reversal in the geometric algebra,
which is analogous to taking the complex conjugate of a complex number.
The reverse $\reverse{q}$ of a quaternion $q$ is the quaternion obtained
by flipping the sign of all 3 imaginary components,
\begin{equation}
\label{quatrev}
  \reverse{q}
  \equiv
  \bigl\{
  \begin{array}{c@{\ \ }c@{\ \ }c@{\ \ }c}
  w & -x & -y & -z \\
  \end{array}
  \bigr\}
  \ .
\end{equation}
The modulus $| q |$ of a quaternion $q$ is,
similarly to the modulus of a complex number,
the square root of the product of itself with its reverse,
\begin{equation}
\label{quatmod}
  | q |
  =
  \sqrt{q \reverse{q}}
  =
  \sqrt{\reverse{q} q}
  =
  \sqrt{w^2 + x^2 + y^2 + z^2}
  \ .
\end{equation}
The modulus $| q |$ of a quaternion is unchanged by any rotation.

One of the advantages of quaternions is that quaternions are easy to invert:
the inverse of a quaternion $q$ is,
again similarly to the inverse of a complex number,
its reverse $\reverse{q}$ divided by its squared modulus,
\begin{equation}
\label{quatinv}
  q^{-1}
  =
  {\reverse{q} \over \reverse{q} q}
  \ .
\end{equation}

The positions of objects in a 3d scene are described by vectors
in the 3d geometric algebra.
According to the array~(\ref{basis3e}) of basis multivectors,
a basis vector $\bgamma_a$ in the 3d geometric algebra
equals minus the pseudoscalar times a bivector,
$\bgamma_a = - I_3 ( I_3 \bgamma_a )$,
which implies that a vector in 3d rotates in the same way as a bivector.
The result is that a vector in a 3d program can be handled as though
it were a bivector, that is, a quaternion with zero scalar component.
The 3-vector position $\brr$ of an object in a 3d scene
can be represented by a quaternion
\begin{equation}
\label{quatvec}
  \brr
  =
  \qi x + \qj y + \qk z
  =
  \bigl\{
  \begin{array}{c@{\ \ }c@{\ \ }c@{\ \ }c}
  0 & x & y & z \\
  \end{array}
  \bigr\}
  \ .
\end{equation}

A 3d rotation is represented by a quaternion of unit modulus,
a rotor in the 3d geometric algebra.
For example, the rotor $R$ corresponding to
a right-handed rotation by angle $\theta$ about the $x$-direction is
\begin{align}
\label{quatrotx}
  R
  &=
  \ee^{- \qi \theta/2}
  =
  \cos(\theta/2) - \qi \sin(\theta/2)
\nonumber
\\
  &=
  \bigl\{
  \begin{array}{c@{\ \ }c@{\ \ }c@{\ \ }c}
  \cos(\theta/2) & - \sin(\theta/2) & 0 & 0 \\
  \end{array}
  \bigr\}
  \ .
\end{align}
More generally,
if $\bn = \qi x + \qj y + \qk z$ is a unit vector in the
direction $\{ x , y , z \}$,
then a right-handed rotation by angle $\theta$ about the $\bn$-direction is
represented by the rotor
\begin{equation}
\label{quatrot}
  R
  =
  \ee^{- \bn \theta / 2}
  =
  \cos(\theta/2) - \bn \sin(\theta/2)
  \ .
\end{equation}
A rotation $S$ following a rotation $R$
is just their quaternionic product $S R$.

Rotating a 3-vector $\brr$ by a rotor $R$ is accomplished by taking the
quaternionic product
\begin{equation}
\label{quatvecrot}
  \brr
  \rightarrow
  R \brr \reverse{R}
  \ ,
\end{equation}
a one-line calculation in a language that supports quaternionic operations.

It should be remarked
that the sign conventions above are the standard ones in physics.
Software implementations of quaternions vary in their conventions.


\subsection{Complex quaternions}

Introduce yet another imaginary $I$
which commutes with the 3 quaternionic imaginaries $\qi$, $\qj$, $\qk$,
\begin{equation}
  I^2
  =
  -1
  \ , \quad
  I \qi = \qi I
  \ , \quad
  I \qj = \qj I
  \ , \quad
  I \qk = \qk I
  \ .
\end{equation}
A complex quaternion $q$ is a quaternion with complex coefficients,
$w = w_R + I w_I$, etc.,
\begin{equation}
\label{qcdef}
  q
  =
  w + \qi x + \qj y + \qk z
  \ .
\end{equation}
A complex quaternion $q$ can be stored as the 8-component object
\begin{equation}
\label{quatc}
  q
  =
  q_R + I q_I
  =
  \biggl\{
  \begin{array}{c@{\ \ }c@{\ \ }c@{\ \ }c}
  w_R & x_R & y_R & z_R \\
  w_I & x_I & y_I & z_I
  \end{array}
  \biggr\}
  \ .
\end{equation}
The top line contains the real part $q_R$ of the quaternion $q$,
the bottom line the imaginary part $q_I$.
A complex quaternion can be implemented in software as a pair of quaternions.

The algebra of complex quaternions is that of the even geometric algebra
in 3+1 spacetime dimensions.
The geometric algebra in 3+1 dimensions is generated by the
$1 + 4 + 6 + 4 + 1 = 16 = 2^4$ orthornormal basis multivectors
\begin{equation}
\label{basis4e}
  \begin{array}{c}
    1 \, , \\
    \mbox{1 scalar}
  \end{array}
  \ \ 
  \begin{array}{c}
    \bgamma_t \, , \  \bgamma_a \, , \\
    \mbox{4 vectors}
  \end{array}
  \ \ 
  \begin{array}{c}
    \bgamma_a \bgamma_b \, , \ I \bgamma_a \bgamma_b \, , \\
    \mbox{6 bivectors}
  \end{array}
  \ \ 
  \begin{array}{c}
    I \bgamma_t \, , \ I \bgamma_a \, , \\
    \mbox{4 pseudovectors}
  \end{array}
  \ \ 
  \begin{array}{c}
    I \equiv \bgamma_t \bgamma_x \bgamma_y \bgamma_z \, , \\
    \mbox{1 pseudoscalar}
  \end{array}
\end{equation}
where indices $a, b$ run over the 3 spatial dimensions $x$, $y$, $z$.
The imaginary $I$ in the complex quaternion~(\ref{quatc})
is just the pseudoscalar of the geometric algebra.
The 6 bivectors of the geometric algebra comprise 3 spatial bivectors
$\bgamma_a \bgamma_b$,
along with 3 bivectors
$I \bgamma_a \bgamma_b$
which look like spatial bivectors multiplied by the pseudoscalar.
The geometrica algebra in 3+1 dimensions is precisely that
of Dirac $\gamma$-matrices in relativistic physics,
which accounts for the notation $\bgamma_m$.

The reverse $\reverse{q}$ of the complex quaternion~(\ref{quatc}) is,
similarly to the reverse~(\ref{quatrev}) of a real quaternion,
the quaternion with all quaternionic imaginary components flipped in sign,
\begin{equation}
\label{quatrevq}
  \reverse{q}
  =
  \biggl\{
  \begin{array}{c@{\ \ }c@{\ \ }c@{\ \ }c}
  w_R & - x_R & - y_R & - z_R \\
  w_I & - x_I & - y_I & - z_I
  \end{array}
  \biggr\}
  \ .
\end{equation}
The complex conjugate $q^\qast$ of the quaternion is the quaternion
with all imaginary $I$ components flipped,
\begin{equation}
\label{quatastc}
  q^\qast
  =
  \biggl\{
  \begin{array}{c@{\ \ }c@{\ \ }c@{\ \ }c}
  w_R & x_R & y_R & z_R \\
  - w_I & - x_I & - y_I & - z_I
  \end{array}
  \biggr\}
  \ .
\end{equation}
The operations of reversal and complex conjugation commute.

The modulus $|q|$ of a complex quaternion is defined,
similarly to the modulus~(\ref{quatmod}) of a real quaternion,
to be the square root of the product $q \reverse{q}$ of
the complex quaternion with its reverse,
but whereas the modulus of a real quaternion is a real number,
the modulus of a complex quaternion is a complex (with respect to $I$) number,
\begin{equation}
  | q |
  \equiv
  \sqrt{q \reverse{q}}
  =
  \sqrt{\reverse{q} q}
  =
  \sqrt{w^2 + x^2 + y^2 + z^2}
  \ .
\end{equation}
The modulus $|q|$ is complex (with respect to $I$) because the components
$w$, $x$, $y$, $z$ are complex.
To obtain a real number,
the absolute value $\lVert q \rVert$ of the complex quaternion $q$,
it is necessary to take the absolute value of the complex modulus,
\begin{equation}
  \lVert q \rVert
  \equiv
  \sqrt{| q | | q |^\ast}
  =
  \bigl\lvert \sqrt{w^2 + x^2 + y^2 + z^2} \bigr\rvert
  \ .
\end{equation}

\subsection{Lorentz transformations}

A Lorentz transformation, a special relativistic rotation of spacetime,
can be represented as a complex quaternion of unit modulus,
a Lorentz rotor $R$, satisfying the unimodular condition
\begin{equation}
\label{quatcunimod}
  R \reverse{R} = 1
  \ .
\end{equation}
The inverse of a Lorentz rotor $R$ is its reverse $\reverse{R}$.
The unimodular condition~(\ref{quatcunimod}) is a complex (with respect to $I$)
condition, which removes 2 degrees of freedom from the 8 degrees of freedom
of a complex quaternion,
leaving the Lorentz group with 6 degrees of freedom, which is as it should be.

Spatial rotations correspond to real unimodular quaternions,
and account for 3 of the 6 degrees of freedom of Lorentz transformations.
For example, a spatial rotation
by angle $\theta$ right-handedly about the $x$-axis
is the real Lorentz rotor
\begin{equation}
  R
  =
  \ee^{- \qi \theta / 2}
  =
  \cos(\theta/2) - \qi \sin(\theta/2)
  \ ,
\end{equation}
or, stored as a complex quaternion,
\begin{equation}
\label{quatcrotx}
  R
  =
  \biggl\{
  \begin{array}{c@{\ \ }c@{\ \ }c@{\ \ }c}
  \cos(\theta/2) & - \sin(\theta/2) & 0 & 0 \\
  0 & 0 & 0 & 0
  \end{array}
  \biggr\}
  \ .
\end{equation}
The expression~(\ref{quatcrotx})
coincides with the earlier expression~(\ref{quatrotx})
for a spatial rotation as a real quaternion, which is as it should be.

Lorentz boosts account for the remaining 3
of the 6 degrees of freedom of Lorentz transformations.
A Lorentz boost is mathematically equivalent to a rotation
by an imaginary angle $I \theta$.
Physicists call the boost angle $\theta$ of a Lorentz boost its rapidity.
The velocity $v$ of the Lorentz boost is the hyperbolic tangent of
the boost angle $\theta$,
\begin{equation}
  v = \tanh\theta
  \ .
\end{equation}
The Lorentz gamma-factor $\gamma$ and associated momentum factor $\gamma v$
that appear in Figure~\ref{beam}
are related to the boost angle $\theta$ by
\begin{equation}
  \gamma = \cosh\theta
  \ , \quad
  \gamma v = \sinh\theta
  \ .
\end{equation}

A Lorentz boost by boost angle $\theta$
along, for example, the $x$-axis is the complex Lorentz rotor
\begin{equation}
  R
  =
  \ee^{- I \qi \theta / 2}
  =
  \cosh(\theta/2) - I \qi \sinh(\theta/2)
  \ ,
\end{equation}
or, stored as a complex quaternion,
\begin{equation}
\label{quatcboostx}
  R
  =
  \biggl\{
  \begin{array}{c@{\ \ }c@{\ \ }c@{\ \ }c}
  \cosh(\theta/2) & 0 & 0 & 0 \\
   0 & - \sinh(\theta/2) & 0 & 0
  \end{array}
  \biggr\}
  \ .
\end{equation}
More generally,
a Lorentz boost by boost angle $\theta$ along a unit 3-vector quaternion
direction $\bn$ is
\begin{equation}
\label{quatcboost}
  R
  =
  \ee^{- I \bn \theta / 2}
  =
  \cosh(\theta/2) - I \bn \sinh(\theta/2)
  \ .
\end{equation}

The most general Lorentz transformation is a Lorentz rotor $R$,
a unimodular complex quaternion.
Any such Lorentz rotor can be decomposed into a
combination of a spatial rotation and a Lorentz boost,
but it is not necessary to carry out such a decomposition.
The arithmetic of complex quaternions takes care of itself.
The rule for composing Lorentz transformations is the same as
the rule for composing spatial rotations:
a Lorentz transformation $S$ following a Lorentz transformation $R$
is just the product
$S R$
of the corresponding complex quaternions.

A general Lorentz rotor $R$ can be written as the exponential
of a complex quaternion angle $\btheta$,
which is itself the product of a complex (with respect to $I$) angle $\theta$
equal to the modulus of $\btheta$,
and a unimodular complex 3-direction $\bn$,
\begin{equation}
\label{rotorc}
  R
  =
  \ee^{- \btheta}
  =
  \ee^{- \theta \bn}
  \ ,
\end{equation}
with
\begin{equation}
\label{anglec}
  \theta
  =
  |\btheta|
  =
  \theta_R + I \theta_I
  \ , \quad
  \bn
  =
  \bn_R + I \bn_I
  \ , \quad
  \reverse{\bn} \bn
  = 1
  \ .
\end{equation}
The complex quaternion 3-direction $\bn$ is a 6-component object,
but the unimodular condition on $\bn$ removes 2 of the degrees of freedom,
leaving $\bn$ with 4 degrees of freedom.
Those 4 degrees of freedom add to the 2 degrees of freedom of the complex
angle $\theta$ to give 6 degrees of freedom,
which is the correct number of degrees of freedom of Lorentz transformations.

\subsection{4-vectors}
\label{4vec-sec}

The positions of objects in a relativistic scene are described by 4-vectors
in the geometric algebra in 3+1 dimensions.
In much the same way that 3-vectors in 3d can be packaged as quaternions,
equation~(\ref{quatvec}),
4-vectors in 3+1 dimensions can be packaged as complex quaternions.
In any geometric algebra,
an odd element of the geometric algebra can be written uniquely as the
product of some particular vector (an odd element)
and an even element of the algebra.
It is convenient to choose the chosen particular vector to be the time
basis vector $\bgamma_t$.
In 3+1 dimensions, an odd element of the algebra is a sum of a vector
$\brr_1$ and a pseudovector $I \brr_2$.
Their sum, the odd multivector $\brr$,
can be written as the product of the time vector $\bgamma_t$
and an element of the even algebra, a complex quaternion $q$,
\begin{equation}
\label{rq}
  \brr
  \equiv
  \brr_1 + I \brr_2
  =
  \bgamma_t q
  \ .
\end{equation}
The odd multivector $\brr$ Lorentz transforms under a Lorentz rotor $R$ as
\begin{equation}
\label{rLorentz}
  \brr
  \rightarrow
  R \bgamma_t q \reverse{R}
  =
  \bgamma_t
  R^\qast
  q \reverse{R}
  \ .
\end{equation}
The conjugated rotor $R^\qast$ appears
in the transformation~(\ref{rLorentz}) because
commuting the time vector $\bgamma_t$ through the rotor $R$
converts the latter to its complex (with respect to $I$) conjugate,
which is true because the time vector commutes with the quaternionic imaginaries
$\qi$, $\qj$, $\qk$,
but anticommutes with the pseudoscalar $I$.
Equation~(\ref{rLorentz}) shows that
the complex quaternion $q$ equivalent to the odd multivector $\brr$
Lorentz transforms as
\begin{equation}
\label{quatcposrot}
  q
  \rightarrow
  R^\qast q \reverse{R}
  \ .
\end{equation}

When you are looking at a scene,
it is light that carries the image of the scene to your eyes.
The perceived position of an object traces back
along the energy-momentum 4-vector
of the light that the object emits and that you see.
The arrowed lightrays in Figure~\ref{beam}
represent the energy-momentum 4-vectors of the lightrays.
When you change your velocity through the scene,
the energy-momentum 4-vectors of the lightrays Lorentz-transform accordingly.
Light travels at light speed,
so the components of the energy-momentum 4-vector $\bp$ of a lightray
(a photon) take the form
\begin{equation}
  \bp = E \{ 1 , \bv \}
\end{equation}
in which the magnitude of the velocity $\bv$ is one, the speed of light,
$v \equiv |\bv| = 1$.
The energy of the lightray (photon) is $E$, and its 3-momentum is $E \bv$.
The energy-momentum 4-vector of an object moving at the speed of light is null;
in the language of the geometric algebra, the energy-momentum 4-vector $\bp$
of a lightray satisfies the null condition
\begin{equation}
  \bp \reverse{\bp} = 0
  \ .
\end{equation}

The positions $\brr$ of objects in a relativistic scene are thus not just
any old 4-vectors, but rather they are null 4-vectors,
satisfying the null condition
\begin{equation}
  \brr \reverse{\brr} = 0
  \ .
\end{equation}
The complex quaternion $q$ equivalent to the 4-vector $\brr$,
equation~(\ref{rq}),
satisfies the null condition
\begin{equation}
  q \reverse{q} = 0
  \ .
\end{equation}

%

In relativistic visualization,
one starts by defining a rest frame,
a frame with respect to which (most) objects are not moving.
Relative to an observer at rest in the scene,
the 3-vector of an object in the scene
at distance $r$ and unit 3-direction $\bn$ is $r \bn$.
The direction $\bn$ to the object is opposite in sign to the direction $\bv$
of a lightray from the object to the observer.
The null 4-vector position $\brr$
of the object is encoded in the null complex quaternion
\begin{equation}
\label{quatcpos}
  q
  =
  r ( 1 + I \bn )
  \ .
\end{equation}
For example, the null 4-vector position of an object lying in the $x$-direction
relative to the observer is
\begin{equation}
\label{quatcposx}
  q
  =
  r \,
  \biggl\{
  \begin{array}{c@{\ \ }c@{\ \ }c@{\ \ }c}
  1 & 0 & 0 & 0 \\
  0 & 1 & 0 & 0
  \end{array}
  \biggr\}
  \ .
\end{equation}
More generally,
if the 3-vector position of the object is
$r \bn = \{ x , y , z \}$,
then the components of the corresponding
null complex quaternion position $q$ are
\begin{equation}
\label{quatcposxyz}
  q
  =
  \biggl\{
  \begin{array}{c@{\ \ }c@{\ \ }c@{\ \ }c}
  r & 0 & 0 & 0 \\
  0 & x & y & z
  \end{array}
  \biggr\}
  \ .
\end{equation}

When the observer rotates spatially or accelerates,
the scene appears to the observer Lorentz-transformed by a Lorentz rotor $R$,
which transforms perceived null complex quaternion positions $q$
according to equation~(\ref{quatcposrot}).
This is similar to the transformation~(\ref{quatvecrot})
of a 3d position by a spatial rotation,
except that the rotor $R^\qast$
in the Lorentz transformation~(\ref{quatcposrot})
is complex-conjugated.
A spatial rotation corresponds to a real rotor $R$,
whose complex conjugate is itself,
in which case the Lorentz transformation~(\ref{quatcposrot})
reduces to the spatial transformation~(\ref{quatvecrot}).
The Lorentz transformation~(\ref{quatcposrot}) of the position $q$
instructs to multiply three complex quaternions
$R^\ast$, $q$, and $\reverse{R}$,
a one-line expression in a language that supports complex quaternion
operations.

As an illustration of how the Lorentz transformation~(\ref{quatcposrot})
works, consider the (simple) example
of an object unit distance from the observer in the (say) $x$-direction,
so $q = ( 1 + I \qi )$,
and the Lorentz transformation is by boost angle $\theta$ in the same
$x$-direction,
so the Lorentz rotor is $R = \ee^{- I \qi \theta / 2}$,
equation~(\ref{quatcboostx}).
Then the Lorentz transformation~(\ref{quatcposrot}) is
\begin{align}
  \biggl\{
  \begin{array}{c@{\ \ }c}
  1 & 0 \\
  0 & 1
  \end{array}
  \biggr\}
  &\rightarrow
  \biggl\{
  \begin{array}{c@{\ \ }c}
  \cosh\frac{\theta}{2} & 0 \\
   0 & \sinh\frac{\theta}{2}
  \end{array}
  \biggr\}
  \biggl\{
  \begin{array}{c@{\ \ }c}
  1 & 0 \\
  0 & 1
  \end{array}
  \biggr\}
  \biggl\{
  \begin{array}{c@{\ \ }c}
  \cosh\frac{\theta}{2} & 0 \\
   0 & \sinh\frac{\theta}{2}
  \end{array}
  \biggr\}
\nonumber
\\
  &=
  \ee^{\theta}
  \biggl\{
  \begin{array}{c@{\ \ }c}
  1 & 0 \\
  0 & 1
  \end{array}
  \biggr\}
  \ ,
\end{align}
which says that the Lorentz-boosted position appears farther away
by the factor $\ee^{\theta}$.
The exponential factor $\ee^{\theta}$ agrees with the conventional
special relativistic Doppler shift formula
\begin{equation}
  \ee^{\theta}
  =
  \cosh\theta + \sinh\theta
  =
  \gamma ( 1 + v )
  =
  \sqrt{1 + v \over 1 - v}
  \ .
\end{equation}
The result also agrees with Figure~\ref{beam},
which illustrates that the scene directly ahead appears farther away
and blueshifted by the Doppler factor $\gamma ( 1 + v )$.

Lorentz transforming a scene typically involves a large number of
Lorentz transformations, one for each point on the scene.
A complex quaternion has 8 components,
and two position vectors $\brr_1$ and $\brr_2$
can be transformed for the price of one by treating one of them, $\brr_2$,
as a pseudovector, and packing the two into a single complex quaternion,
\begin{equation}
\label{quatcpos2}
  q
  =
  r_1 ( 1 + I \bn_1 )
  -
  I
  r_2 ( 1 + I \bn_2 )
  \ .
\end{equation}
If the 3-vector components of the two positions are
$r_i \bn_i = \{ x_i , y_i , z_i \}$, $i = 1, 2$,
then the components of the packed complex quaternion $q$ are
\begin{equation}
  q
  =
  \biggl\{
  \begin{array}{c@{\ \ }c@{\ \ }c@{\ \ }c}
  r_1 & x_2 & y_2 & z_2 \\
  - r_2 & x_1 & y_1 & z_1
  \end{array}
  \biggr\}
  \ .
\end{equation}
The same Lorentz transformation formula~(\ref{quatcposrot}) applies.

\section{Conclusion}

In the future, children will learn the mysteries of
special and general relativity by playing with relativistic flight simulators.
Relativity will cease to be something accessible only to the cognoscenti.

The first task in coding a relativistic flight simulator
is to implement Lorentz transformations.
The right way to do that is with complex quaternions.
This paper shows how.


\bibliographystyle{splncs04}

\end{document}